# Direct quantification of brown algae-derived fucoidans in human plasma by a fluorescent probe assay

U. Warttinger[1], C. Giese[1], J. Harenberg[2], Roland Krämer[1]

Correspondence to:
Roland Krämer, phone 0049 6221 548438, fax 0049 6221 548599
E-mail: kraemer@aci.uni-heidelberg.de

1 Heidelberg University, Inorganic Chemistry Institute, Im Neuenheimer Feld 270, 60129 Heidelberg, Germany
2 Heidelberg University, Medical Faculty Mannheim, Maybachstr. 14, 68169 Mannheim, Germany

**Abstract**

Fucoidan is a generic term for a class of fucose rich, structurally diverse sulfated polysaccharides that are found in brown algae and other marine organisms. Depending on the species from which the fucoidan is extracted, a wide variety of biological activities including antitumor, antiinflammatory, immune-modulating, antiviral, antibacterial and pro- and anticoagulant activities has been described. Fucoidans have the advantage of low toxicity and oral bioavailibiity and are viable drug candidates, preclinical and pilot clinical trials show promising results. The availability of robust assays, in particular for analysing the blood levels of fucoidan, is a fundamental requirement for pharmacokinetic analysis in drug development projects. This contribution describes the application of a commercially availbale, protein-free fluorescent probe assay (Heparin Red) for the direct quantification of several fucoidans (from *Fucus vesiculosus*, *Macrocystis pyrifera*, and *Undaria pinnatifida*) in human plasma. By only minor adapation of the established protocol for heparin detection, a concentration range 0,5 – 20 µg/mL fucoidan can be addressed. A preliminary analysis of matrix effects suggests acceptable interindividual variability and no interference by endogeneous chondroitin sulfate. This study identifies the Heparin Red assay as a simple, time-saving mix-and-read method for the quantification of fucoidans in human plasma.

**Keywords**

Fucoidan, assay, fluorescence, plasma, Heparin Red

**Introduction**

Fucoidans are structurally diverse, fucose-rich sulfated polysaccharides that are commonly extracted from brown algae. A major structural motif are repeating, sulfated fucose units, linked by α-1,3 or α-1,4 bonds (scheme 1). [1] In contrast to glycosaminoglycans, most fucoidans are highly branched.

Fucoidans are a natural bioactive ingredients of marine functional foods. Depending on the species from which they have been extracted, fucoidans exhibit a broad spectrum of biological activities, including antitumor, anti-inflammatory, immuno-modulating, antiviral, antibacterial, anticoagulant and procoagulant effects [2-14]

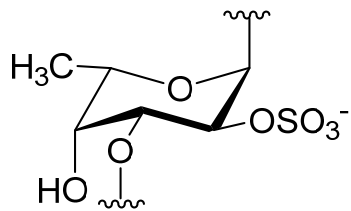

**Scheme 1**. α-1,3-linked, sulfated fucose unit, as present in many fucoidans. Sulfation pattern is variable, fucose may be monosulfated, disulfated or non-sulfated.

Currently, there are no approved uses for fucoidan in medical applications. It is, however, considered a promising drug candidate due to lack of toxicity, good biocompatibility, oral bioavailability, and encouraging results in preclinical and sporadic early-stage clinical trials. For drug development and approval, pharmacokinetic analysis needs to be considered. A fundamental requirement for understanding the pharmacokinetics of fucoidan is the availability of robust analytical methods, in particular for determining the blood levels of the target compound.

A limited number of assays for the detection of fucoidan in plasma or serum samples has been reported. A competitive ELISA was developed in 2005 and applied to the quantification of an *Undaria pinnatifida* fucoidan in plasma after oral ingestion of the compound by human volunteers. Plasma levels in the range 4 - 13 µg/mL have been reported. [15] Later, a more sensitive (low ng/mL range) sandwich Elisa was described for *Cladosiphon okanuramus* fucoidan in human serum and urine; no cross reactivity to heparin was observed.[16,17] Both ELISAs require sample pretreatment (boiling and centrifugation), and several washing steps and overall incubation times of several hours are necessary. A polyion-sensitive membrane electrode for the direct, potentiometric detection of fucoidan in buffer in the range 0,5 – 50 µg/mL was recently described; fucoidan could also be indirectly measured in fetal bovine serum (2,5 – 7,5 µg/mL), using a modified method that includes protamine titration of the fucoidan and detection of free protamine by a protamine-sensitive membrane electrode.[18] To our knowledge, none of these approaches has been translated in a generally available commercial assay.

Besides these assays for plasma and serum levels of fucoidan, a size exclusion chromatography HPLC method with UV detection at 210 nm was applied to fucoidan detection in buffer, as well as in urine at concentrations in the range about 5-250 µg/mL. [19] A filter paper spot test using methylene blue staining allows the colorimetric detection of fucoidan in buffer and bovine serum albumin solutions; the method requires an analyte quantity of about 1 µg [20]. Very recently, a sensitive fluorescence assay was reported for fucoidan detection in aqueous extracts, using the nucleic acid stain SYBR gold [21], and a modified toluidine blue assay specific for sulfated polysaccharides, including fucoidan, in the presence of carboxylated polysaccharides. [22]

We describe here the quantification of Fucoidan in spiked human plasma samples by the commercially available Heparin Red Kit. The latter is a fluorescent probe assay that has been developed for the direct detection of heparin in plasma samples.[23, 24] Heparins display certain structural similarities to fucoidans with respect to their polymeric (polysaccharide) nature and high negative charge density due to sulfation. The 615 nm fluorescence emission of Heparin Red lies at the edge of the "optical transparency window" of blood at longer wavelengths, so that interference by colored components and autofluorescence is minimised. The molecular probe is a polyamine-functionalized perylene diimide that forms a supramolecular complex with the polyanionic polysaccharide target. Aggregation of the probe molecules at the polysaccharide template results in contact quenching of fluorescence (scheme 2). Since this response mechanism is independent of any anticoagulant activity, the Heparin Red assay is also capable of detecting non-anticoagulant heparins [25] and is emerging [26] as a valuable tool for pharmacokinetic studies of this promising class of drug candidates since the standard heparin assays based on interaction with coagulation factors are not applicable.

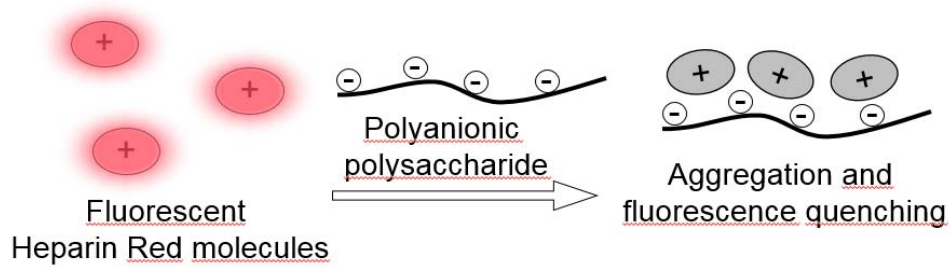

**Scheme 2.** Schematic representation of fluorescence quenching of the molecular probe Heparin Red in the presence of polyanionic polysaccharides.

Here we describe for the first time the application of the Heparin Red Kit to the detection of several fucoidans in spiked human plasma samples, by implementing only a minor modification of the routine protocol for heparin detection.

## Materials and Methods

### Instrumentation

*Fluorescence measurements*

Fluorescence was measured with a microplate reader Biotek Synergy Mx (Biotek Instruments, Winooski, VT, USA), excitation at 570 nm, emission recorded at 605 nm, spectral band width 13,5 nm, read height of 8 mm.

*Microplates*

For fluorescence measurements 96 well microplates, polystyrene, Item No 655076, were purchased from Greiner Bio-One GmbH, Frickenhausen.

*Pipettes*

Transferpette® 100-1000µl, Transferpette® 10-100µl, Transferpette®-8 20-200µl, purchased from Brand GmbH, Wertheim. Rainin Pipettes 100-1000µl, 20-200µl, and 2-20µl purchased from Mettler Toledo, OH, USA.

*Data analysis*

Data were analyzed using Excel (Microsoft Office 10) and OriginPro 8.1. First order exponential decay was applied to Heparin Red calibration curves with *F. vesiculosus* fucoidan. Fucoidan concentrations of the single-donor plasmas shown in table 3 are given relative to calibration curves for the respective analyte in spiked pooled normal plasma.

### Reagents

*Kits*

The Heparin Red® Kit was a gift from Redprobes UG, Münster, Germany. Kit components: Heparin Red solution, Product No HR001, Lot 01-003, and Enhancer Solution, Product No ES001, Lot 004.

*Fucoidans*

Fucoidans were purchased from Sigma-Aldrich GmbH, Steinheim: Fucoidan from *Fucus vesiculosus*, purity >95%, product number F8190, Lot SLBN8754V. Fucoidan from *Macrocystis pyrifera*, purity > 85%, product number F8065, Lot SLBP2754V. Fucoidan from *Undaria pinnatifida*, purity > 95%, product number F8315, Lot SLBQ0431V.

*Plasma*

Pooled normal plasma (Lot A1161, Cryocheck, Precision Biologic, NS, Kanada) consists of a pool of normal citrated human plasma from a minimum of 20 healthy donors. Storage conditions of the pooled normal plasma prior to spiking with fucoidans was -80°C. The fucoidan spiked plasma samples were stored at -20°C.

Five single-donor citrated plasmas of healthy individuals were provided by the Blood Bank of the Institute for Clinical Transfusion Medicine and Cell Therapy Heidelberg (IKTZ). The single donor plasmas were applied as matrix for the detections of table 3. The fucoidan spiked single donor plasmas were stored at -20°C.

.

*Other*

Aqueous solutions were prepared with HPLC grade water purchased from VWR, product No 23595.328. Chondroitinsulfate A sodium salt from bovine trachea, product number C9819, Lot SLBQ0017V, 85% pure according to 13C NMR spectrum (Certificate of analysis), and Chondroitinase ABC from *Proteus vulgaris*, lyophilized powder, product number C3667, were received from Sigma-Aldrich GmbH, Steinheim.

**Assays**

*Heparin Red® Kit*

For determination of fucoidan concentrations in plasma samples, the protocol of the provider for a 96-well microplate assay was followed, with minor adaptations in order to address different concentration ranges of fucoidan. Mixtures of Heparin Red solution and Enhancer solution were freshly prepared as follows:

| Enhancer solution | Heparin Red solution | Fucoidan concentration range | sensitivity |
|---|---|---|---|
| 9 mL | 100 µL | 2-20 µg/mL | low |
| 9 mL | 50 µL | 1-10 µg/mL | medium |
| 9 mL | 25 µL | 0,5-5 µg/mL | high |

20 µL of the fucoidan spiked plasma sample was pipetted into a microplate well, followed by 80 µL of the Heparin Red – Enhancer mixture. The microplate was introduced in the fluorescence reader and mixing was performed using the plate shaking function of the microplate reader (setting "high", 3 minutes). Immediately after mixing, fluorescence was recorded within 1 minute.

*Spiked plasma samples*

Plasma samples containing defined mass concentrations of fucoidans were prepared by adding aqueous solutions (10 vol%) of the corresponding fucoidan to plasma and vortexing, to achieve a concentration of 100 µg/mL. Fucoidan concentrations required for the detections were adjusted by further dilution of this 100 µg/mL stock solution with the same plasma and vortexing. The spiked plasma samples were stored at -20°C for at least one day and up to 2 months, thawed at room temperature and vortexed before use. Similarly, a chondroitinsulfate plasma spike of concentration 48 µg/mL was prepared.

*Chondroitin sulfate assay*

For determination of chondroitin sulfate in plasma samples with the Heparin Red Kit, the protocol of the provider for a 96-well microplate assay was followed with minor adaptations, using a mixture of 9 mL Enhancer solution and 100 µL Heparin Red solution. Chondroitin sulfate spiked plasma samples were used instead of heparin spiked plasma samples. Sample pre-treatment with chondroitinase ABC was performed at 0,27 U/mL enzyme concentration in the plasma sample for 20 minutes.

## Results and discussion

**Detection of fucoidans in the concentration range 1-10 µg/mL**

The Heparin Red fluorescence assay was applied to three commercially available fucoidans, extracted from different seaweed species: *Fucus vesiculosus*, *Macrocystis pyrifera*, and *Undaria pinnatifida* . Literature data for selected structural features of these types of fucoidan are given in table 1.

| Fucoidan | MW (kDa) | Fucose % | Uronic acid % | Sulfation ratio |
|---|---|---|---|---|
| *F. vesiculosus* | 82,5 | 43,1 | 8,7 | 0,81 |
| *M. pyriferia* | 176,4 | 30,5 | 12,7 | 1,1 |
| *U. pinnatifida* | 51,7 | 32,6 | 4,0 | 0,85 |

**Table 1**. Structural properties of the fucoidan types used in this study, as reported in [9]. MW refers to the peak average molar weight as determined by gel permeation chromatography. % refers to absolute mass per cent. Sulfation ratio is the average number of sulfate groups per monosaccharide moiety (note that fucoidans also contain other monosaccharides such as galactose, not included in the present table).

Data in table 1 indicate a similar sulfation degree between 0,81 and 1,1 for the three fucoidan types. The sulfation degree is of particular relevance in this context since it determines the negative charge density, and electrostatic interactions contribute significantly to the strong binding of the polycationic molecular probe Heparin Red [23, 24]. An additional, although minor contribution to negative charge density of fucoidans may arise from negatively charged carboxylate groups of uronic acids.

The response of the Heparin Red assay to fucoidan plasma spikes in the concentration range 1-10 µg/mL is shown in figure 1. Since Heparin Red is expected to form non-fluorescent aggregates with fucoidan, the fluorescence intensity dereases with increasing plasma levels of the analyte. Response to the different fuciodans is similar, slightly weaker for the *M. pyrifera* compound. In view of the limited purity of the commercial fucoidan preparations (95% for *F. vesiculosus* and *U. pinnatifida*, 85% for *M. pyrifera*), these subtle differences in response behaviour will not be discussed here with respect to the different charge densities of the fucoidans (table 1). The results in Fig. 1 suggest that the Heparin Red assay is suitable for the direct detection of various fucoidans in human plasma in the concentration range about 1-10 µg/mL.

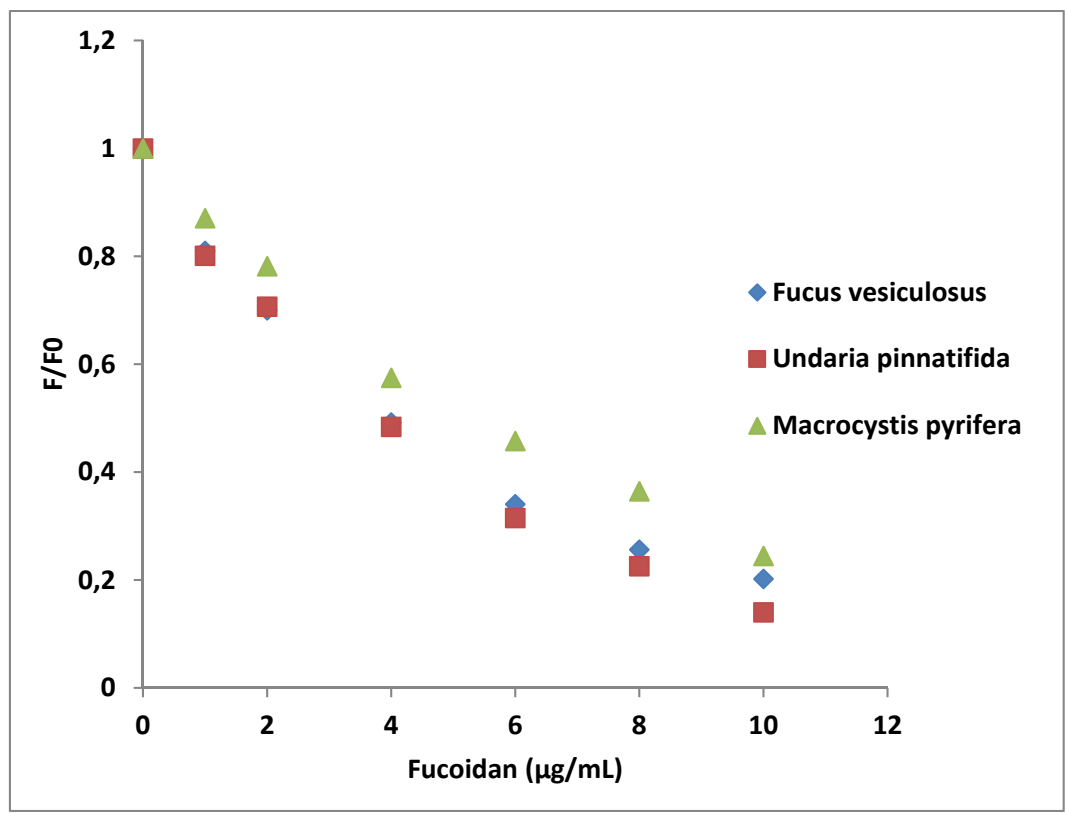

**Figure 1.** Normalized dose response curves of the Heparin Red fluorescence assay in the presence of various fucoidans (*F. vesiculosus*, *M. pyriferia*, *U. pinnatifida*). Spiked pooled normal plasma samples. Manually performed microplate assay, using a mixture of 50 µL Heparin red and 9 mL Enahncer solution. Excitation at 570 nm, fluorescence emission at 605 nm. Averages of duplicate determinations; CVs (averaged over all concentrations) between 1,9 (*M. pyrifera*) and 6,5 % (*F. vesiculosus* and *U. pinnatifida*), depending on the fucoidan type.

**Tuning the sensitivity of the Heparin Red assay for fucoidan**

For a selected fucoidan (*Fucus vesiculosus*), we have extended the sensitive range by variation of the fluorescent probe concentration in the detection solution (see Materials and Methods for details). Half concentration of the probe results in a stronger relative fluorescence quenching at lower fucoidan plasma levels, and a sensitive range of about 0,5 – 5 µg/mL can be addressed. At double concentration of the probe, the detection range is shifted to higher concentrations, about 2-20 µg/mL. (Figure 2).

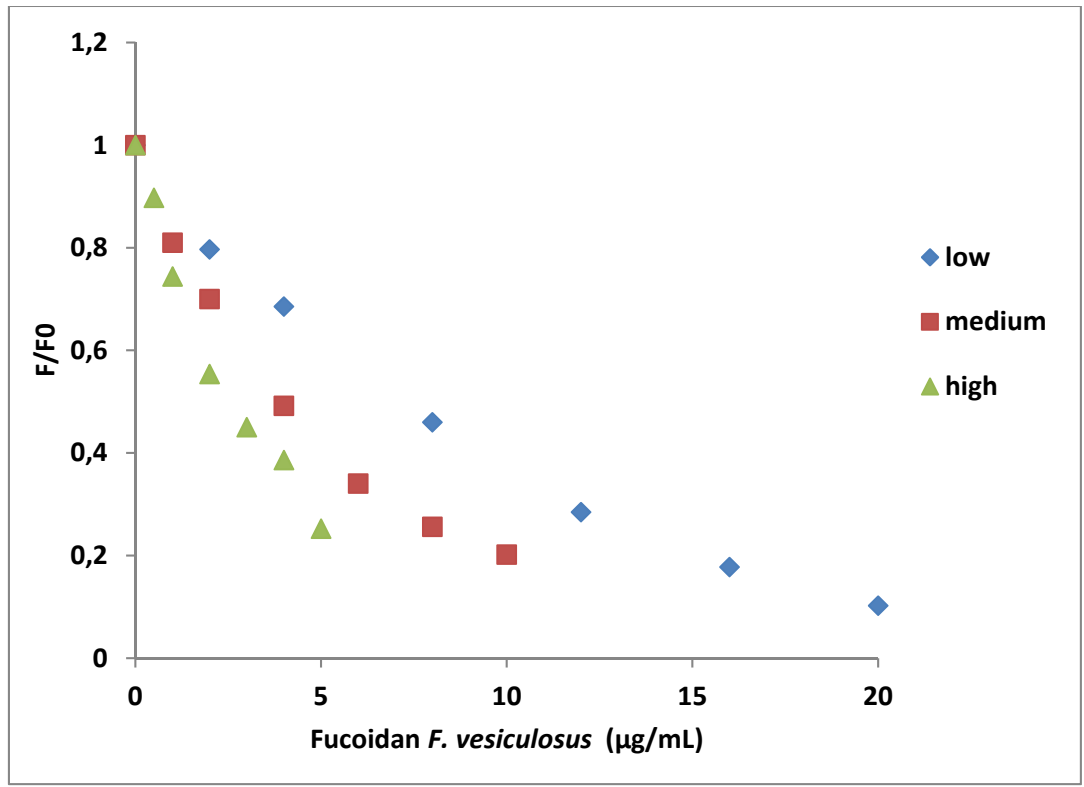

**Figure 2.** Normalized dose response curves of the Heparin Red fluorescence assay in the presence of *F. vesiculosus* fucoidan. Sensitivity of the Heparin red assay (low, medium, high) is tuned by the concentration of the fluorescent probe in the detection solution (see Materials and Methods for details). Spiked pooled normal plasma samples. Manually performed microplate assay Excitation at 570 nm, fluorescence emission at 605 nm. Averages of triplicate determinations; CVs (averaged over all concentrations) between 2,0 and 7,9 %, depending on sensitivity level (low: 7,9%; medium 2,0 %; high: 3,2%).

The detection medium for low sensitivity (blue diamonds in Fig. 2) corresponds to that of the standard protocol for heparin detection. Compared with unfractionated heparin [23], a higher mass concentration of fucoidan is required to trigger 50% fluorescence quenching. This is interpreted with the higher charge density of heparin, -1,7 vs about -0,81 considered for *F. vesiculosus* fucoidan (sulfation degree, table 1), and the tendency of Heparin Red to form charge neutral aggregates with the polyanionic target.[24]

Standard deviations (n=8) for selected fucoidan concentrations 8, 4 and 2 µg/mL that yield about 50% fluorescence quenching for assay sensitivities low, medium and high, respectively, are given in table 2. In this evaluation, intraassay-variability is somewhat higher for the highest sensitivity level.

|  | Low: c = 8 µg/mL | Medium: c = 4 µg/mL | High: c = 2 µg/mL |
|---|---|---|---|
| $F/F_0$ average | 0,46 | 0,52 | 0,60 |
| $F/F_0$ range | 0,41 - 0,49 | 0,49 - 0,58 | 0,51 - 0,68 |
| SD | 0,03 | 0,03 | 0,05 |

**Table 2.** Intraassay variability (n=8) of *F. vesiculosus* fucoidan detection by the Heparin Red assay at the different sensitivity levels (low, medium, high, compare figure 2).

**Plasma matrix effects**

Interference of fucoidan determination in plasma by heparin is obvious, in particular since heparin triggers stronger response of Heparin Red than fucoidan does. For a subgroup of people under heparin therapy, the Heparin Red kit may therefore not provide accurate results in plasma fucoidan determination.

We have also addressed the question of a potential interference by endogeneous glycosaminoglycans (GAGs). Chondroitin sulfate (CS) is considered the main glycosaminoglycan in human plasma, and several studies report an average CS mass concentrations of about 5 µg/mL in normal human plasma [27-29] Also, several studies describe that most of the plasma CS is an undersulfated form, with sulfation degrees between 25% and 54% per disaccharide, what corresponds to a charge density of – 0,62 to - 0,77 per monosaccharide moiety. [27, 29, 30] This is lower than the charge density of the investigated fucoidans.

We have also tested potential CS interference experimentally by pre-treating the pooled normal plasma used in this study with chondroitinase ABC. This enzyme cleaves CS into shorter oligosaccharide chains and ultimately disaccharides that are not detected by Heparin Red. If any interfering chondroitin sulfate is present in the plasma sample, enzymatic pre-treatment should result in a higher fluorescence of Heparin Red relative to untreated plasma since the oligosaccharide chains that trigger fluorescence quenching have been eliminated. We see, however, no difference in fluorescence intensity between the chondroitase pre-treated and control sample after incubation with Heparin Red (Fig. 3, left). As a control experiment, the plasma was spiked with 48 µg/mL CS from bovine lung. The sample without enzymatic pre-treatment displays a significantly lower fluorescence, indicating a weak response of Heparin Red to this CS (Fig. 3, right). The chondroitinase treated sample, on the other hand, has a fluorescence comparable to that of a CS free sample, indicating effective depolymerization of CS to short saccharides that do not trigger a response of Heparin Red. We conclude that there is no significant interference by chondroitin sulfate in normal plasma of healthy individuals. In this context, it would be interesting to apply the Heparin Red Kit for monitoring the potential activity of fucoidanases [31] in human plasma, in order to verify detection of fucoidans if the presence of interfering glycosaminoglycans such as heparin is suspected.

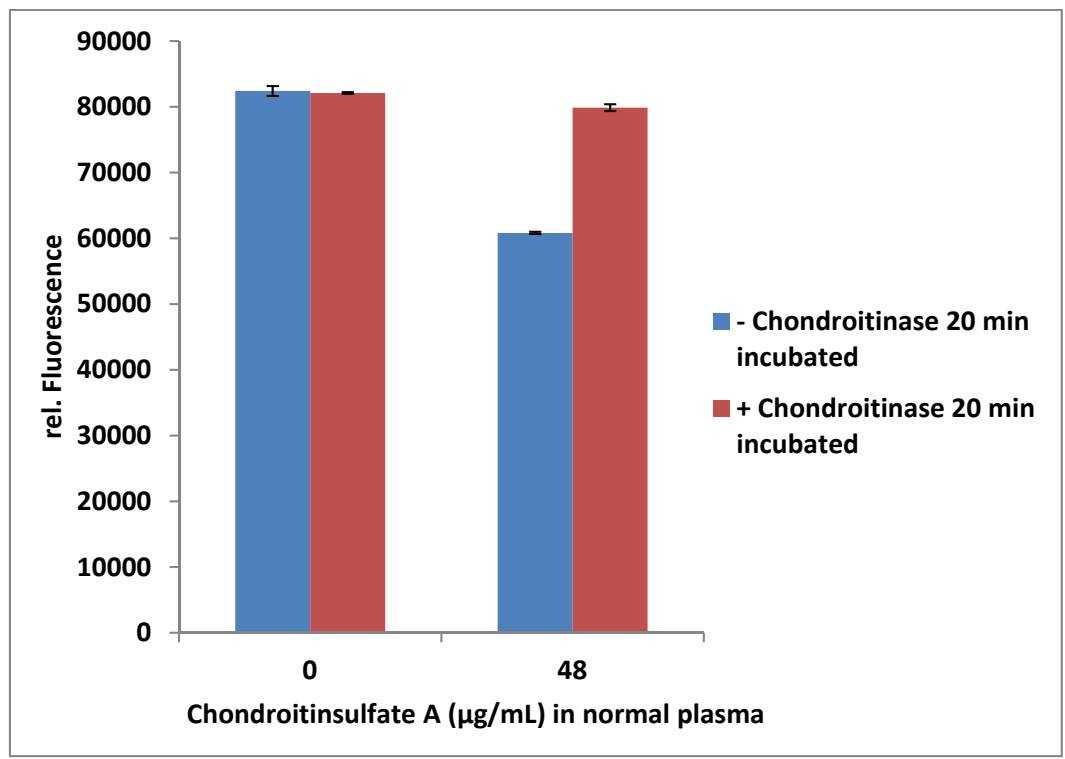

**Figure 3.** Fluorescence response of the Heparin Red fluorescence assay to unspiked pooled normal plasma without (blue bars) or with (red bars) pre-treatment by chondroitinase (0,27 U/mL), and to chondroitin sulfate (48 µg/mL) spiked pooled normal plasma without (blue bars) or with (red bars) pre-treatment by chondroitinase. Manually performed microplate assay. Excitation at 570 nm, fluorescence emission at 605 nm. Averages of duplicate determinations.

The interindividual variability of the fucoidan determination by Heparin Red was explored using *F. vesiculosus* fucoidan spiked plasma samples of five healthy donors. Fucoidan concentrations were assigned by comparison with a calibration curve (corresponding to the red curve in figure 3) with spiked pooled normal plasma. Interindividual variation (table 3) is within an acceptable range, although higher than the intraassay variability for repeated measurements of the same plasma sample (table 2), indicating a substantial influence of the plasma matrix on assay results.

| Spike conc. µg/mL | 0,0 | 1,0 | 2,0 | 4,0 | 6,0 | 8,0 |
| --- | --- | --- | --- | --- | --- | --- |
| Average (µg/mL) | -0,1 | 0,9 | 1,8 | 4,3 | 6,5 | 8,1 |
| Range (µg/mL) | -0,3 - 0,2 | 0,7 - 1,2 | 1,6 - 2,1 | 3,8 - 4,8 | 5,6 - 7,1 | 7,8 – 8,6 |
| SD (µg/mL) | 0,19 | 0,18 | 0,17 | 0,38 | 0,58 | 0,30 |

**Table 3**. Determination of *F. vesiculosus* fucoidan concentrations (µg/mL) by the Heparin Red assay in spiked plasmas of healthy donors (n=5). Concentrations were determined relative to a calibration curve with spiked pooled normal plasma. Averages of duplicate determinations.

## Conclusion

This study describes the application of a direct fluorescence assay (Heparin Red) for the quantification of fucoidans in human plasma. Although the assay has been optimized for heparin detection, it can be applied to fucoidans with only minor adapations of the routine protocol to achieve sensitive detection in the range 0,5 - 20 µg/mL with <10% intraassay variability. In comparison with other literature described analysis methods, we see a major strength of this microplate assay in the simple, time-saving mix-and-read protocol (less than 10 minutes for measuring 16 samples in duplicate, calibration included). In contrast, ELISAs [15-17] for specific fucoidans require plasma or serum sample pre-treatment, and several washing and incubation steps, resulting in overall assay times of several hours. Analysis time of a potentiometric membrane method for fucoidan detection in a bovine serum sample [18] is also significantly higher since based on a protamine titration with multiple additions. A limitation of the Heparin Red assay is its lower sensitivity when compared with one ELISA [16] (low ng/mL range), and the cross-reactivity with the anticoagulant drug heparin. Heparin interference may [15] or may not [16] be present with ELISAs, and has not been addressed for the potentiomtric membrane method. Regarding endogeneous glycosaminoglycans, we have shown that interference of fucoidan detection with Heparin Red by chondroitin sulfate, the major plasma glycosaminoglycan, is unlikely in normal plasma. Fucoidan concentrations in spiked plasmas of healthy donors (n=5) were determined relative to a calibration curve with pooled normal plasma, with acceptable variability between individuals.

We see great potential of the Heparin Red assay for pharmacokinetic and clinical studies in fucoidan related drug development projects and treatment studies of diseases. Also, the subject of oral bioavailability of fucoidans still requires considerably more understanding [9], and the assay seems suitable for the screening of fucoidan plasma levels after oral ingestion of fucoidans or fucoidan-rich marine functional foods. In addition,